\documentclass[twocolumn]{jpsj2} 
\title{Modulation of the Shubnikov-de Haas Oscillation in Unidirectional Lateral Superlattices}

\author{Akira \textsc{Endo}\thanks{E-mail address: akrendo@issp.u-tokyo.ac.jp} and Yasuhiro \textsc{Iye}}

\inst{Institute for Solid State Physics, University of Tokyo, 5-1-5 Kashiwanoha, Kashiwa, Chiba 277-8581}

\abst{The amplitude and phase of Shubnikov-de Haas oscillations have been analyzed in detail for two-dimensional electron gases subjected to a weak unidirectional periodic potential modulation. The amplitude is suppressed, accompanied by inversion of the phase, at the maximum bandwidth conditions at low magnetic fields. The suppression is gradually taken over by the enhancement with the increase of the magnetic fields. The suppression and the enhancement are attributable to the collisional and the diffusion contribution of the modulated potential to the conductivity, respectively, the former (the latter) being dominant at low (high) magnetic fields. A theoretical calculation that takes the two types of contributions into account shows semi-quantitative agreement with experimental traces.}

\kword{lateral superlattice, Shubnikov-de Haas oscillation, commensurability oscillation}

\begin{document}
\maketitle

\section{Introduction} 
The Shubnikov-de Haas oscillation (SdHO) in a two-dimensional electron gas (2DEG) is a manifestation of Landau quantization, $E_N = (N+1/2) \hbar \omega_\mathrm{c}$, in the magnetoresistance \cite{Ishihara86}. The oscillation is periodic in $1/B$, and damps with decreasing magnetic fields since the cyclotron energy $\hbar \omega_\mathrm{c}=\hbar eB/m^*$ (with $m^*$ the electron effective mass) diminishes with respect to both the thermal blurring $k_\mathrm{B}T$ of the Fermi energy $E_\mathrm{F}$ and the disorder broadening $\Gamma$ of the Landau levels (LLs). In a unidirectional lateral superlattice (ULSL), a 2DEG subjected to a one-dimensional (1D) periodic modulation, another analogous oscillation periodic in $1/B$ emerges. The oscillation, known as the commensurability oscillation (CO) \cite{Weiss89,Gerhardts89,Winkler89}, results from the commensurability between the cyclotron radius $R_\mathrm{c}=\hbar k_\mathrm{F}/eB$ and the period $a$ of the modulation, and also damps with decreasing $B$. Here $k_\mathrm{F}=\sqrt{2\pi n_e}$ represents the Fermi wave number with $n_e$ the areal density of electrons. In the CO, the energy difference between adjacent flat-band conditions (see below), $(a k_\mathrm{F}/2)\hbar \omega_\mathrm{c}$, takes the place of $\hbar \omega_\mathrm{c}$ in the SdHO. Since $(a k_\mathrm{F}/2)\gg 1$ in most experiments performed so far, the SdHO vanishes and only the CO survives at the low end of magnetic fields. At low enough temperatures where thermal damping is not so severe, however, the two classes of oscillation coexist over a range of magnetic field with an intriguing interplay. The behavior of the SdHO under the influence of the simultaneously present CO is the main subject of the present study.

A unidirectional periodic modulation of the electrostatic potential,
\begin{equation}
 V(x) = V_0 \cos(qx)
\label{potmod}
\end{equation}
with $q = 2 \pi/a$, lifts the degeneracy of the LLs. The energy becomes dependent on the position $x_0$ of the guiding center and reads, in the first order perturbation theory valid for a weak modulation $V_0 \ll E_\mathrm{F}$,
\begin{equation}
E_{N,qx_0} = \left(N + \frac{1}{2}\right)\hbar\omega _\mathrm{c}  + V_{N,B} \cos(qx_0),
\label{Energy}
\end{equation}
where $V_{N,B}=V_0 e^{ - u/2} L_N(u)$ with $L_N(u)$ the Laguerre polynomial, $u = q^2 l^2 /2$, and $l = \sqrt{\hbar/eB}$ the magnetic length. $V_{N,B}$ at the Fermi energy can be approximated, using an asymptotic expression for $N \gg 1$ appropriate for low-magnetic-field range relevant to the present study, by,
\begin{equation}
V_B = 
V_0 \sqrt {\frac{2}{\pi q R_\mathrm{c}}} \cos \left(q R_\mathrm{c} - \frac{\pi}{4} \right).
\label{VBapp}
\end{equation}
$V_B$ oscillates with $B$, and the width of the Landau bands $2|V_B|$ takes maximum at
\begin{equation}
\frac{2 R_\mathrm{c}}{a}=n+\frac{1}{4}\hspace{10mm}(n=1,2,3,...),
\label{bandmax}
\end{equation}
and vanishes at the \textit{flat band conditions}
\begin{equation}
\frac{2 R_\mathrm{c}}{a}=n-\frac{1}{4}\hspace{10mm}(n=1,2,3,...).
\label{flatband}
\end{equation}
The oscillation of the Landau bandwidth is the origin of the CO and, at the same time, is responsible for the modulation of the amplitude and the phase of the SdHO.

The modulation of the SdHO amplitude was first reported by Overend \textit{et al}. \cite{OverendG98} They exploited a periodic modulation of magnetic field instead of electrostatic potential. For a magnetic field modulation, eqs. (\ref{bandmax}) and (\ref{flatband}) interchange their roles, with eq. (\ref{bandmax}) representing the flat band conditions. The underlying physics, however, is basically the same for both types of modulations. The authors reported that the SdHO amplitude remains large at the flat band conditions [eq. (\ref{bandmax})], while is suppressed at the maximum bandwidth conditions [eq. (\ref{flatband})]. Further study by the same group \cite{Edmonds01,Shi02} showed the phase inversion of the SdHO at the maximum bandwidth conditions when the bandwidth is larger than $\hbar \omega_\mathrm{c}$. They explained their observation as an effect of the modulated density of states (DOS), which affects the conductivity through the collisional contribution. The phase inversion is essentially the same phenomenon as the even-odd filling-factor switching reported for a ULSL with strong electrostatic potential modulation in the quantum Hall regime \cite{Tornow96}, and is attributed to the van Hove singularities in the DOS\@. 

Similar suppression of the SdHO amplitude at the maximum bandwidth conditions [eq. (\ref{bandmax})] was also reported for electrostatic ULSLs with a strong modulation amplitude induced by a patterned InGaAs stressor layer \cite{Milton00}. Interestingly, the authors observed that the trend is reversed at higher magnetic fields; the amplitude is \textit{enhanced} at the maximum bandwidth conditions. The origin of this enhancement, as well as of the inversion of the tendency with the magnetic field, still lacks comprehensive explanation.

The purpose of the present paper is to achieve more quantitative understanding of the behavior of the SdHO under periodic modulation. In the previous works quoted above \cite{OverendG98,Edmonds01,Shi02,Tornow96,Milton00}, ULSLs with a large amplitude periodic modulation are employed, which is obviously advantageous in attaining the modulation of the SdHO strong enough to be readily observed. However, the amplitude of the periodic modulation or the width of Landau bands exceeding 10\% of the Fermi energy seems to be rather incompatible with the perturbative treatment as in eq.\ (\ref{Energy}). In the present work, we keep $|V_B|/E_\mathrm{F}$ $\leq$ $V_0/E_\mathrm{F}$ to be less than five percent, which validates the comparison of the experimental data with perturbation theories. We employ low temperatures, which allows us to observe the SdHO down to low magnetic fields ($\sim$ 0.05 T) where $V_0 / \hbar \omega_\mathrm{c}$ becomes large even for our small $V_0$.

\section{Experimental}
\begin{figure}[tb]
\includegraphics[bbllx=45,bblly=220,bburx=510,bbury=800,width=8.5cm]{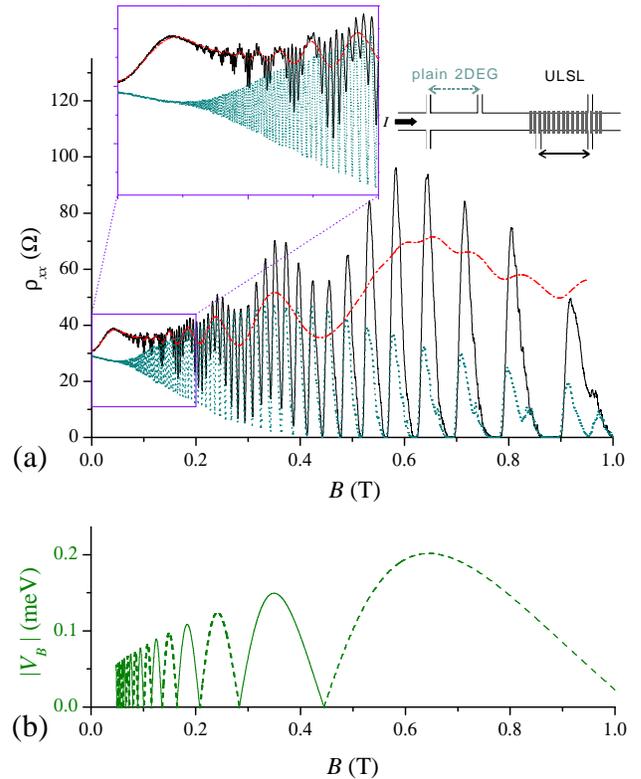}
\caption{(a) (Color online) Magnetoresistance traces for the ULSL (solid line) and the plain 2DEG (dotted line) taken at 15 mK for sample A\@. A trace for the ULSL at 4.2 K is also shown with dot-dashed line \cite{NoteDiffCryo}. The upper right inset depicts the schematic configuration of the sample. (b) The absolute value of $V_B$ [eq.\ (\ref{VBapp})], with $V_B$ $>$ 0 ($<$ 0) plotted by solid (dashed) line.}
\label{rawVB}
\end{figure}

We examine two ULSL samples with slightly different characteristics, as tabulated in Table \ref{Samples}. The two samples are fabricated from two GaAs/AlGaAs 2DEG wafers having nominally the same structure \cite{structure} but differing in the carrier mobilities owing to the conditions of the molecular beam epitaxy (MBE) chamber used for the growth. The electrostatic potential modulation is introduced via strain-induced piezoelectric effect \cite{Skuras97} by placing negative electron-beam (EB) resist on the surface \cite{Endo00E}. As depicted in the right inset of Fig.\ \ref{rawVB} (a), each sample is patterned into a Hall bar that has two sets of voltage probes to measure the longitudinal and Hall resistivity of the section with (ULSL) and without (plain 2DEG) the potential modulation at the same time. The values of the mobility $\mu$ and the density $n_e$ tabulated in Table \ref{Samples} are those measured after brief illumination by LED, the condition in which the magnetoresistance traces presented in this paper are taken. No sign of deterioration in $\mu$ or change in $n_e$ arising from the microfabrication process is discerned.

\begin{table}
\caption{Sample parameters}
\begin{center}
\begin{tabular}{lcc}
\hline
  & Sample A & Sample B \\
\hline
Period $a$ (nm) & 231 & 184 \\
Amplitude $V_0$ (meV) & 0.39 & 0.31 \\
Mobility $\mu$ (m$^2$/Vs) & 69 & 101 \\
Quantum mobility $\mu_\mathrm{Q}$ (m$^2$/Vs) & 11.9 & 7.2 \\
Density $n_e$ (10$^{15}$ m$^{-2}$) & 2.9 & 2.9 \\
\hline
\end{tabular}
\label{Samples}
\end{center}
\end{table}

For Sample B, the periodic modulation is introduced by a EB-resist grating having a conventional \textit{periodic} line-and-space pattern. In this approach, higher harmonics inevitably mixes in the potential profile as the period $a$ becomes large compared to the depth $d$ ($=$ 90 nm for the present samples) of the 2DEG plane from the surface \cite{Endo05HH}. An unconventional strategy of employing a \textit{quasiperiodic} pattern is applied for Sample A with a larger period. Here, slabs of EB-resist are placed on the ``$L$''s of the Fibonacci sequence ``$LSLLSLSL...$'', with $L$=104 nm and $S$=64 nm [$L/S$ is set to $\phi$=$(1+\sqrt{5})/2$=1.618...]. Generally in such Fibonacci ULSLs, the analysis of the CO reveals several frequency components, each corresponding to one of the self-similar generations of a potential profile mutually scaled by the factor $\phi$ \cite{Endo07I,Endo07FCO}. In the particular sample explored in this study (Sample A), a single component with effective period $a$=231 nm, corresponding to the average distance between adjacent ``$S$''s, overwhelmingly dominates the potential profile, allowing the profile to be virtually regarded as periodic. Although somewhat counterintuitive, a simple sinusoidal potential profile described by eq.\ (\ref{potmod}) is realized better in Sample A than in Sample B because of the absence of the higher harmonics. 

\begin{figure}[tb]
\includegraphics[bbllx=50,bblly=80,bburx=550,bbury=800,width=8.5cm]{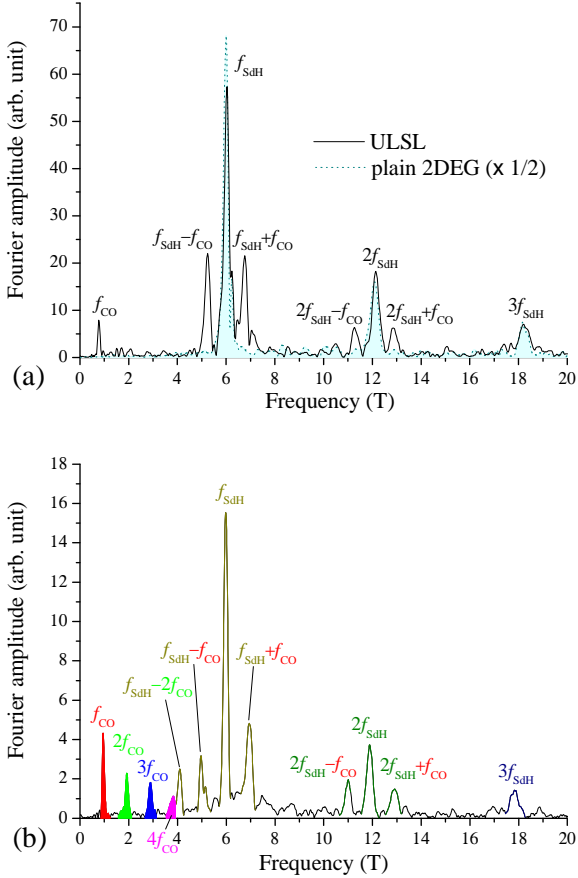}
\caption{(Color online) Fourier transform of $(d^2/dB^2)[\rho_{xx}(B) / \rho_0]$ versus $1/B$ for the ULSL in sample A (a) and B (b). Fourier spectra for the plain 2DEG is also shown for (a) by dotted line with the shade underneath. Peak positions are given by combinations of integer multiples of $f_\mathrm{SdH}$ and $f_\mathrm{CO}$, where $f_\mathrm{SdH}$ ($f_\mathrm{CO}$) represents the fundamental frequency for the SdHO (CO).}
\label{d2BFFT}
\end{figure}

Cursory characterization of the ULSL samples can be done by performing a Fourier transform of the $(d^2/dB^2)[\rho_{xx}(B) / \rho_0]$ vs. $1/B$ curve, where $\rho_{xx}(B)$ represents longitudinal magnetoresistance and $\rho_0=\rho_{xx}(0)$. The second derivative conveniently eliminates the slowly-varying background from the magnetoresistance traces. The Fourier spectra shown in Fig.\ \ref{d2BFFT} exhibit peaks corresponding to the CO ($f_\mathrm{CO}$), the SdHO ($f_\mathrm{SdH}$), and their combinations ($p f_\mathrm{SdH} \pm q f_\mathrm{CO}$ with $p$, $q$ integers). The CO peaks up to the fourth harmonic ($f_\mathrm{CO}$, $2 f_\mathrm{CO}$, $3 f_\mathrm{CO}$, and $4 f_\mathrm{CO}$) are seen for Sample B, indicating the presence of the corresponding harmonic contents in the potential profile. For Sample A, in contrast, only one CO peak is found, justifying the description by eq.\ (\ref{potmod}) of the potential profile. The presence of the peaks at $p f_\mathrm{SdH} \pm q f_\mathrm{CO}$ reveals the interplay between the two types of oscillation that will be discussed in detail in the following sections. The higher harmonics of the CO in Sample B do affect the interplay as indicated by the presence of a peak $f_\mathrm{SdH} - 2 f_\mathrm{CO}$ \cite{highharm}.

More quantitative account of the actual potential profile requires analyses of the CO amplitude. The values of the fundamental component $V_0$ tabulated in Table \ref{Samples} are obtained by such analyses that take the effect of damping by scattering into account \cite{Endo00E}. Higher harmonic contents can also be evaluated by analyses using Fourier band pass filters \cite{Endo05HH,Endo07FCO}, and the second, the third, and the fourth harmonic components for Sample B are found to be $V_2$=0.10 meV, $V_3$=0.07 meV, and $V_4$=0.05 meV, respectively. Although these higher harmonics can in principle complicate the analysis below for Sample B, we neglect them for simplicity and assume, in the rest of this paper, that both Sample A and Sample B have a potential profile of the form given by eq.\ (\ref{potmod}).

Measurements are performed in a top-loading dilution fridge at the base temperature $T \simeq$ 15 mK. Standard low-frequency (13 Hz) ac lock-in technique with a current $I_\mathrm{rms}$=10 nA is employed for the  resistivity measurement. We have checked that the electron heating by the current is negligible in the magnetic-field range of the present interest ($|B|$ $<$ 1 T) by reducing $I_\mathrm{rms}$ down to 0.5 nA, for which the  magnetoresistance trace shows no difference except for much worse signal-to-noise ratio. A low sweep rate $dB/dt$=0.01 T/min is employed in order to avoid undesired hysteresis of the superconducting magnet. The remnant hysteresis is further calibrated by using the simultaneously measured Hall resistivity.

\section{Results and Discussion}
\subsection{Experimentally obtained Shubnikov-de Haas oscillation}
Figure \ref{rawVB} (a) shows magnetoresistance traces of Sample A for both the ULSL and the adjacent plain 2DEG\@. Magnetoresistance of the ULSL at 4.2 K is also shown \cite{NoteDiffCryo}, which essentially represents the pure CO for $|B|$ $<$ $\sim$0.5 T where the SdHO has already damped out. In the low magnetic field range ($|B|$ $<$ $\sim$ 0.25 T), the SdHO of the ULSL is suppressed at the maxima of the CO, while remains almost unaltered from that of the plain 2DEG at the minima of the CO. In contrast, the SdHO amplitude is observed to be enhanced at the CO maxima for higher magnetic fields. This is most clearly seen for the CO peak at 0.64 T\@. In Fig.\ \ref{rawVB} (b), the half width of Landau bands calculated by eq.\ (\ref{VBapp}) are plotted. It can readily be confirmed that, as is well known \cite{Gerhardts89,Winkler89}, the maxima and the minima of the CO correspond to the maximum bandwidth and the flat band conditions, respectively. 

\begin{fullfigure}[tb]
\includegraphics[bbllx=20,bblly=290,bburx=810,bbury=700,width=18cm]{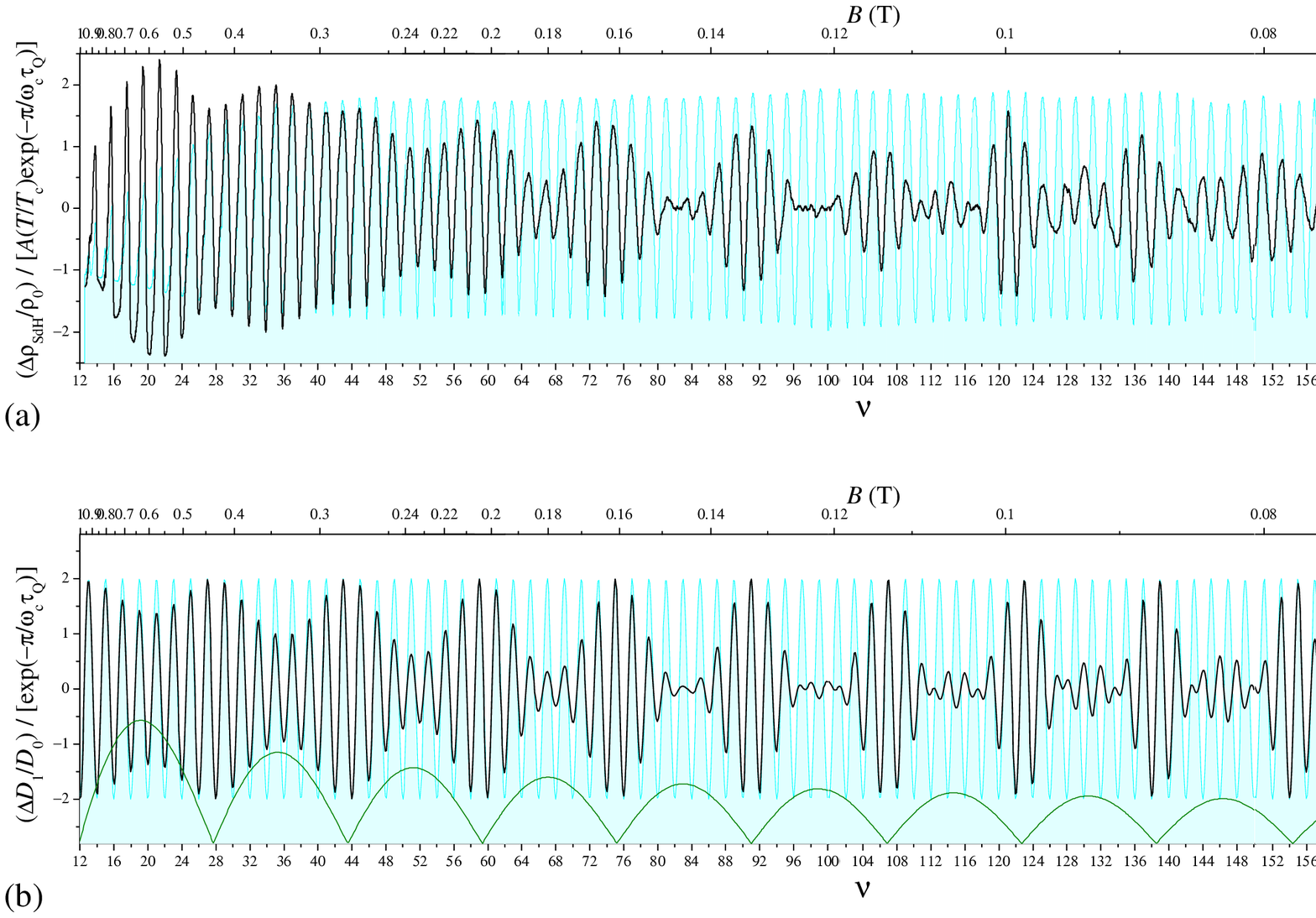}
\caption{(Color online) (a) Experimentally obtained SdHO, divided by the thermal and the scattering damping factors, for sample A\@. See text for details. (b) Density of states divided by the exponential damping factor, calculated by eq.\ (\ref{D1plain}) or eq.\ (\ref{D1ULSL}) with $V_0$ = 0.49 meV. The thin line with the shade underneath and the thick line represent the plain 2DEG and the ULSL, respectively, plotted against the filling factors (bottom axes) or magnetic fields (top axes). The half width of the Landau bands, $|V_B|$, is also plotted in (b) (right axis).}
\label{Sim}
\end{fullfigure}

To look into the details of the behavior of the SdHO, the rapidly oscillating parts of the magnetoresistance, $\Delta \rho_\mathrm{SdH} / \rho_0$, are extracted and plotted against the Landau level filling factor $\nu$=$n_e h/eB$ in Fig.\ \ref{Sim} (a) for both the ULSL and the plain 2DEG of Sample A\@. The corresponding magnetic field is shown on the top axis. The extraction of the $\Delta \rho_\mathrm{SdH} / \rho_0$ is done by applying a Fourier high pass filter to the $\rho_{xx} / \rho_0$ vs. $1/B$ curve, with the threshold set at a frequency higher than $f_\mathrm{CO}$, and further by subtracting the average of the upper and lower envelopes, as was done in the analysis of the CO \cite{Endo00E}. The SdHO is known to damp with decreasing magnetic field by the thermal damping factor $A(T/T_\mathrm{c})$ and the scattering damping factor $\exp (-\pi / \omega_\mathrm{c} \tau_\mathrm{Q})$, where $A(x)$=$x/\sinh(x)$, $k_\mathrm{B} T_\mathrm{c}$=$\hbar \omega_\mathrm{c} / 2 \pi^2$, and $\tau_\mathrm{Q}$ the single particle (quantum) scattering time. Traces shown in Fig.\ \ref{Sim} are $\Delta \rho_\mathrm{SdH} / \rho_0$ divided by these damping factors, applying to both traces the value of the quantum mobility $\mu_\mathrm{Q} = e \tau_\mathrm{Q} / m^*$=11.9 m$^2$/Vs determined from the actual damping of the SdHO of the plain 2DEG\@. The factor $A(T/T_\mathrm{c})$ turns out to barely deviate from unity, reflecting the fact that the thermal damping is negligibly small at this low temperature. Since spins are unresolved for the magnetic-field range in the present study, the minima and maxima of the SdHO are expected to take place at even and odd filling factors, respectively. This is actually what we observe for the plain 2DEG\@. For the ULSL, the SdHO amplitudes are suppressed at the maximum bandwidth conditions, $B$=0.239, 0.183, 0.148, 0.125, 0.107, 0.094, 0.084, 0.076, 0.069, 0.064 T [$n$=3, 4,...,12 in eq.\ (\ref{bandmax}), note the $|V_B|$ plotted in Fig.\ \ref{Sim} (b)]. The SdHO amplitude at the suppressed conditions decrease with decreasing magnetic fields until it vanishes at $B$ $\sim$ 0.125 T, and then revives at still lower magnetic fields but with the position of peaks and dips inverted; there, the minima and maxima occurs at odd and even filling factors, respectively. This is the even-odd transition reported in refs.\ \citen{Edmonds01,Shi02,Tornow96,Milton00}, which was attributed, by using numerically calculated DOS and the resultant conductivity, to the broadening and the van Hove singularity of individual LLs that result in the peaks of DOS at even filling factors. 

\subsection{Comparison with calculated density of states \label{cmpDOS}}
It is well established that the $\Delta \rho_\mathrm{SdH} / [\rho_0 A(T/T_\mathrm{c})]$ is proportional to the oscillatory part of the DOS at the Fermi energy, $\Delta D (E_\mathrm{F}) / D_0$, at low temperatures provided the $|\Delta \rho_\mathrm{SdH} / \rho_0 A|$ is not too large \cite{Ishihara86,Coleridge89}. In this subsection, we make a detailed comparison of the experimentally obtained SdHO described in the previous subsection with a calculated DOS\@.

First we deduce an analytic expression for the DOS under a periodic modulation eq.\ (\ref{potmod}), instead of resorting to a numerical calculation as was done in the previous works. We start by recalling the DOS for a plain 2DEG \cite{Ishihara86}. The disorder broadened line shape of each LL peak is approximated here for simplicity by a Lorentzian, $P(E) = (\Gamma/\pi)/(E^2+\Gamma^2)$, with the width $\Gamma$ independent of $B$. The DOS is obtained by summing up the LL peaks, including the factor 2 for spin degeneracy, 
\begin{eqnarray}
\displaystyle{ D(E)} & = & \displaystyle{ \frac{2}{2\pi l^2}\sum\limits_{N=0}^\infty {P(E - E_N)}} \hspace{30mm} \nonumber \\
 & = & \displaystyle{ D_0 \left\{ 1 + 2\sum\limits_{k = 1}^\infty  \cos \left[ 2\pi k\left( \varepsilon - \frac{1}{2} \right) \right]e^{ - 2\pi k\gamma} \right\}}, \nonumber \\
\label{DOSplain}
\end{eqnarray}
where $D_0 = m^* / \pi \hbar^2$ represents the constant DOS of a 2DEG in the absence of magnetic field. We introduced dimensionless parameters, $\varepsilon = E / \hbar \omega _\mathrm{c}$ and $\gamma = \Gamma / \hbar \omega _\mathrm{c}$. In the second line of eq.\ (\ref{DOSplain}), we made use of the Poisson sum formula \cite{approxminus}. Since $\exp(-2 \pi \gamma) \ll 1$ for small magnetic fields, it is usually a good approximation to keep only the $k$=1 term in the summation: $D(E) \simeq D_0 + \Delta D_1(E)$ with
\begin{equation}
\frac{\Delta D_1(E)}{D_0} = -2 \cos(2 \pi \varepsilon) \exp(-2 \pi \gamma).
\label{D1plain}
\end{equation}
The proportionality $\Delta \rho_\mathrm{SdH} / \rho_0 A \propto \Delta D_1 / D_0$ implies $\Gamma = \hbar / 2 \tau_\mathrm{Q}$ by the comparison of the exponential factors. Therefore the quantum mobility $\mu_\mathrm{Q} =$ 11.9 m$^2$/Vs corresponds to $\Gamma =$ 0.073 meV\@.

In Fig.\ \ref{Sim} (b), we plot the oscillatory part of the DOS at $E_\mathrm{F} = n_e / D_0$ \cite{constEF} divided by the damping factor, $\Delta D_1(E_\mathrm{F}) / [D_0 \exp(-2 \pi \gamma)] = -2 \cos(\varepsilon_\mathrm{F})$ with $\varepsilon_\mathrm{F} = E_\mathrm{F} / \hbar \omega_\mathrm{c} = \nu / 2$. Comparison of Figs. \ref{Sim} (a) and (b) confirms that the relation $\Delta \rho_\mathrm{SdH} / [\rho_0 A(T/T_\mathrm{c})] \propto \Delta D_1 (E_\mathrm{F}) / D_0$ actually holds for our plain 2DEG, aside from the deviation at higher magnetic-field range $|B|$ $>$ 0.4 T where the peak height of $\Delta \rho_\mathrm{SdH} / [\rho_0 A(T/T_\mathrm{c})]$ diminishes accompanying the onset of spin-splitting, which was not taken into consideration in the calculation of the DOS\@. 

Upon switching on the potential modulation eq.\ (\ref{potmod}), each LL peak further broadens from $P(E)$ to (with $t = q x_0$)
\begin{equation}
P(E,V_B) = \frac{1}{\pi} \int_0^\pi P(E - V_B \cos t)dt.
\end{equation}
This can be formally rewritten as
\begin{eqnarray}
\displaystyle{ \frac{i}{2 \pi} \left( \frac{1}{\sqrt{E-V_B+i\Gamma}\sqrt{E+V_B+i\Gamma}} \right.} \hspace{15mm} \nonumber \\
\displaystyle{ \left. - \frac{1}{\sqrt{E-V_B-i\Gamma}\sqrt{E+V_B-i\Gamma}} \right),} \nonumber
\end{eqnarray}
and displays broadening or, when $V_B$ is large compared to $\Gamma$, splitting (van Hove singularities) of a LL peak. Correspondingly, the DOS becomes
\begin{equation}
\displaystyle{ D(E,V_B) = \frac{2}{2\pi l^2}\sum\limits_{N=0}^\infty {P(E - E_N,V_B)}} \hspace{30mm} \nonumber
\end{equation}
\begin{eqnarray}
\displaystyle{ = D_0 \left\{ 1 + 2\sum\limits_{k = 1}^\infty \frac{1}{\pi} \int_0^\pi \! \! \! \cos \left[ 2\pi k\left( \varepsilon - v_B \cos t - \frac{1}{2} \right) \right] dt \right. } \nonumber \\
\displaystyle{ \Biggl. \times e^{ - 2\pi k\gamma} \Biggr\}}
\nonumber
\end{eqnarray}
\begin{equation}
= D_0\left\{ 1 + 2\sum\limits_{k = 1}^\infty  { \cos \left[ 2\pi k\left( \varepsilon - \frac{1}{2} \right) \right]J_0 \left( 2\pi k v_B \right)e^{ -2\pi k \gamma} } \right\},
\label{DOSULSL}
\end{equation}
with $v_B = V_B / \hbar \omega _\mathrm{c}$, and $J_0(x)$ the Bessel function of order zero. $\Delta D_1(E)$ acquires an extra factor $J_0(2 \pi v_B)$:
\begin{equation}
\frac{\Delta D_1(E)}{D_0} = -2 J_0(2 \pi v_B) \cos(2 \pi \varepsilon) \exp(-2 \pi \gamma).
\label{D1ULSL}
\end{equation}

\begin{figure}[t]
\includegraphics[bbllx=30,bblly=70,bburx=560,bbury=250,width=8.5cm]{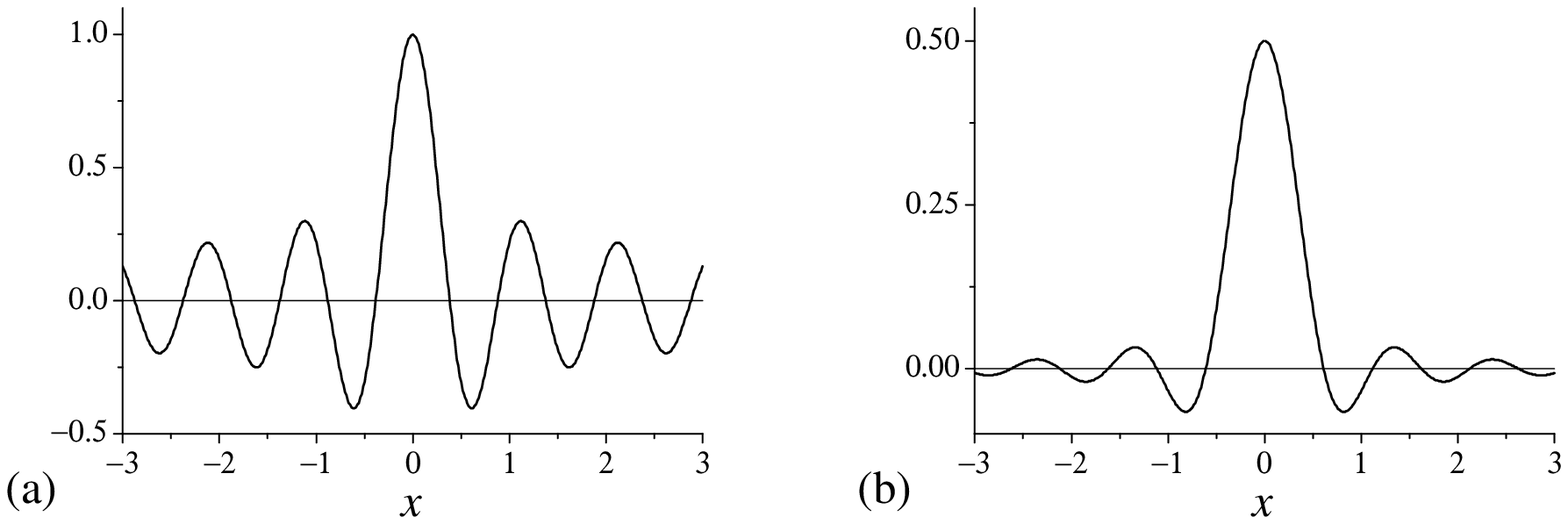}
\caption{Plots of $J_0(2 \pi x)$ (a) and $J_1(2 \pi x)/(2 \pi x)$ (b).}
\label{Bessels}
\end{figure}

With the decrease of $B$, $v_B$ oscillates periodically in $1/B$ back and forth around $v_B = 0$, increasing its amplitude proportionally to $B^{-1/2}$ [see eq.\ (\ref{VBapp})]. As shown in Fig.\ \ref{Bessels} (a), the function $J_0(2 \pi v_B)$ decrease from 1 with the increase of $|v_B|$, becomes zero at $|v_B|$ = 0.3827... $\simeq$ 3/8 and then changes its sign. Therefore the oscillation of $\Delta D_1$ takes minimum amplitude at the maximum bandwidth conditions while $|v_B|$ stays less than 3/8, disappears when a maximum of $|v_B|$ touches $\sim$3/8, reappears with inverted sign for $|v_B|$ larger than 3/8 \cite{stilllarge}. Thus the position where the oscillation of $\Delta \rho_\mathrm{SdH} / [\rho_0 A(T/T_\mathrm{c})]$ vanishes is the landmark of $|V_B| = 0.3827 \hbar \omega_\mathrm{c}$ \cite{evenodd}, on an assumption that the relation $\Delta \rho_\mathrm{SdH} / [\rho_0 A(T/T_\mathrm{c})] \propto \Delta D_1 (E_\mathrm{F}) / D_0$ holds also for ULSLs. From the position we can determine the modulation amplitude $V_0$ using eq.\ (\ref{VBapp}). The oscillation disappears and a small sign of inverted peak is observed at $B$ = 0.125 T in Fig.\ \ref{Sim} (a). From this we obtain $V_0$ = 0.49 meV\@. The DOS calculated using $V_0$ = 0.49 meV plotted in Fig.\ \ref{Sim} (b) reproduce the line shape of the observed $\Delta \rho_\mathrm{SdH} / [\rho_0 A(T/T_\mathrm{c})]$ in Fig.\ \ref{Sim} (a) quite well for $B < \sim$ 0.25 T, confirming the proportionality relation. The value $V_0$ = 0.49 meV is roughly 25\% larger than that deduced from the CO, the reason of this discrepancy remains to be elucidated.

For Sample B, the modulation amplitude is not large enough to achieve the collapse or the peak/dip inversion of the SdHO\@. However, the modulated SdHO trace similar to Fig.\ \ref{Sim} (a) is well reproduced for $B < \sim$ 0.2 T by eq.\ (\ref{D1ULSL}) with $V_0$ = 0.38 meV, again some 20\% larger than the value determined from the CO\@.

For higher magnetic fields, SdHO amplitude is enhanced at the maximum bandwidth conditions, as mentioned earlier. This is obviously not reproduced in the calculated DOS and requires an alternative explanation. We note in passing that the onset of the spin-splitting for the ULSL shifts to a higher magnetic field due to the detrimental effect of the modulation-induced broadening of LLs on the Zeeman gap, and is outside the magnetic-field range for the present study.

\subsection{Comparison with calculated conductivities}

\subsubsection{Analytic expressions for conductivities at low temperatures}
It has been pointed out by Peeters and Vasilopoulos that a periodic potential modulation $V(x)$ alters the conductivity of a 2DEG via two different routes \cite{Peeters92}. \textit{The collisional (hopping) contribution} corresponds to the effect of the DOS we have just discussed above. Peaks in the DOS boost the conductivity hence the resistivity, through the increase in the scattering rate. The other route, \textit{the diffusion (band) contribution}, results from the drift velocity,
\begin{equation}
v_y = \frac{1}{\hbar}\frac{\partial E_{N,(-q l^2 k_y)}}{\partial k_y } = \frac{q V_B}{eB} \sin(q x_0),
\end{equation}
which enhances the $\sigma_{yy}$ hence the $\rho_{xx}$. Here, we made use of the relation $x_0 = -l^2 k_y$. It is this effect that mainly contributes to the CO, with $v_y^2$ being maximum (zero) at the maximum-bandwidth (flat-band) conditions. It is worth pointing out that the diffusion contribution has no counterpart in a plain 2DEG and therefore vanishes with $V_0 \rightarrow 0$, while the collisional contribution is basically an ordinary SdH effect with a due modification introduced by the periodic modulation.

Shi \textit{et al}. calculated by a perturbation theory the collisional ($\sigma_{xx}^\mathrm{col}=\sigma_{yy}^\mathrm{col}$) and the diffusion conductivities ($\sigma_{yy}^\mathrm{dif}$) for 2DEGs under 1D modulation of both an electrostatic potential and a magnetic field \cite{Shi02}. Taking only the electrostatic potential, eq.\ (\ref{potmod}), relevant to the present study into account, they read,
\begin{eqnarray}
\displaystyle{ \sigma_{xx}^{\rm{col}}  = \frac{e^2}{h}\Gamma ^2 \int_{ - \infty }^\infty  {dE\left( - \frac{\partial f}{\partial E} \right)} \sum\limits_{N = 0}^\infty  {(2N + 1)}} \nonumber \\
\displaystyle{\times \int_0^\pi  {dtP^2 (E - E_{N,t})}},
\label{sgmcol}
\end{eqnarray}
and
\begin{eqnarray}
\displaystyle{ \sigma_{yy}^{\rm{dif}} = \sigma_0 \frac{V_0 ^2}{\hbar \omega_\mathrm{c} (ak_\mathrm{F}/2)^2} 2\pi \int_{ - \infty }^\infty  {dE\left( - \frac{\partial f}{\partial E} \right)}} \hspace{10mm} \nonumber \\
\displaystyle{ \times \sum\limits_{N = 0}^\infty {\left[ e^{ - u/2} L_N (u) \right]^2 } \int_0^\pi  {dtP(E - E_{N,t})\sin ^2 t} },
\label{sgmdif}
\end{eqnarray}
respectively, where $f(E) = \{ 1+\exp[(E-E_\mathrm{F})/k_\mathrm{B}T] \}^{-1}$ is the Fermi-Dirac distribution function and $\sigma_0 = n_e e^2 \tau / m^* = (e^2/h) (E_\mathrm{F} / \Gamma_0)$ represents the conductivity at zero magnetic field with $\tau = \hbar/2 \Gamma_0$ the momentum relaxation time \cite{difdef}.  In eq.\ (\ref{sgmcol}), contributions from inter-Landau-band hoppings are omitted. The authors of ref.\ \citen{Shi02} made comparison with experimental results by numerically evaluating these equations. Here, 
we will take a further step and deduce from eqs. (\ref{sgmcol}) and (\ref{sgmdif}) approximate analytic formulae appropriate for comparison with our experimental data.

Firstly, since we employ a low temperature ($T \simeq$ 15 mK) in our measurement, the derivative of the Fermi-Dirac distribution function can safely be approximated by the delta function $\delta (E-E_\mathrm{F})$, resulting in,
\begin{equation}
\sigma_{xx}^{\rm{col}}  = \frac{e^2 }{h}\Gamma ^2 \sum\limits_{N = 0}^\infty  {(2N + 1)} \int_0^\pi  {dtP^2 (E_\mathrm{F}  - E_{N,t})} 
\label{sgmcolLT}
\end{equation}
and
\begin{eqnarray}
\displaystyle{ \sigma_{yy}^{\rm{dif}} = \sigma_0 \frac{ 2\pi V_0 ^2}{\hbar \omega_\mathrm{c} (ak_\mathrm{F}/2)^2} \sum\limits_{N = 0}^\infty {\left[ e^{ - u/2} L_N (u) \right]^2 }} \nonumber \\
\displaystyle{ \times \int_0^\pi  {dtP(E_\mathrm{F}  - E_{N,t})\sin ^2 t}}.
\label{sgmdifLT}
\end{eqnarray}
Equation (\ref{sgmcolLT}) can be rewritten along the same line as in eq.\ (\ref{DOSplain}). Leaving the detail of the derivation to the Appendix, the calculation leads to, up to the leading term,
\begin{eqnarray}
\displaystyle{ \frac{\sigma_{xx}^{\rm{col}}}{\sigma_0}  =  \frac{\Gamma _0}{\Gamma }\gamma ^2 \left\{ 1 + 2\sum\limits_{k = 1}^\infty  {\left( 1 + 2\pi k \gamma \right)} \right. } \hspace{20mm} \nonumber \\
\displaystyle{ \Biggl. \times 
\cos \left[ 2\pi k\left( \varepsilon _\mathrm{F}  - \frac{1}{2} \right) \right]  J_0 ( 2\pi kv_B )e^{ - 2\pi k\gamma }  \Biggr\} }. \nonumber \\
\label{sgmcolFin}
\end{eqnarray}
In eq.\ (\ref{sgmdifLT}) the term $e^{-u/2} L_N(u)$ may be replaced by its asymptotic expression at the Fermi energy, since $P(E_\mathrm{F} - E_{N,t})$ takes significant value only at $N \sim N_\mathrm{F} = [E_\mathrm{F} / \hbar \omega_\mathrm{c}]$ (the integer part of $E_\mathrm{F} / \hbar \omega_\mathrm{c}$), the index of LL in which the Fermi level resides ($N_\mathrm{F} \gg 1$ for low magnetic fields), and therefore
\begin{eqnarray}
\displaystyle{ \sigma_{yy}^{\rm{dif}} = \sigma_0 \frac{ 2\pi ^2 V_0 ^2}{\hbar \omega _\mathrm{c} (ak_\mathrm{F}/2)^2} {\left[ \sqrt{\frac{2}{\pi q R_\mathrm{c}}}\cos \left( q R_\mathrm{c} -\frac{\pi}{4} \right) \right]^2}  } \nonumber \\
\displaystyle{ \times \frac{1}{\pi }\int_0^\pi  {dt \sin ^2 t\sum\limits_{N = 0}^\infty  {P(E_\mathrm{F}  - E_{N,t})} } }.
\end{eqnarray}
As we have done in eq.\ (\ref{DOSULSL}), we employ eq.\ (\ref{DOSplain}) to rewrite the summation $\sum\nolimits_{N = 0}^\infty  {P(E_\mathrm{F}  - E_{N,t})}$, and use relations $(1/\pi) \int_0^\pi {\sin^2 t} dt = 1/2$, $(1/\pi) \int_0^\pi {\cos (x \cos t) \sin^2 t} dt = J_1(x)/x$, and $(1/\pi) \int_0^\pi {\sin (x \cos t) \sin^2 t} dt = 0$, to finally obtain
\begin{eqnarray}
\displaystyle{ \frac{\sigma_{yy}^{\rm{dif}}}{\sigma_0} = \frac{V_0 ^2}{\hbar \omega _\mathrm{c} E_\mathrm{F} (ak_\mathrm{F}/2)}\cos ^2 \left( qR_\mathrm{c}  - \frac{\pi }{4} \right) \Biggl\{ 1 +\Biggr.  } \hspace{15mm} \nonumber \\
\displaystyle{ \left.  4\sum\limits_{k = 1}^\infty  {\cos \left[ {2\pi k\left( \varepsilon _\mathrm{F}  - \frac{1}{2} \right)} \right]} \frac{J_1 (2\pi k v_B)}{2\pi k v_B}e^{ - 2\pi k\gamma }  \right\} }, \nonumber \\
\label{sgmdifFin}
\end{eqnarray}
where $J_1(x)$ is the Bessel function of order one. The conductivities are translated to resistivities by the inversion of the conductivity tensor. For not too small magnetic fields ($B > \sim$0.05 T), $\rho_{xx}^\mathrm{col} / \rho_0 \simeq (\omega_\mathrm{c} \tau)^2 \sigma_{xx}^\mathrm{col} / \sigma_0$ and $\rho_{xx}^\mathrm{dif} / \rho_0 \simeq (\omega_\mathrm{c} \tau)^2 \sigma_{yy}^\mathrm{dif} / \sigma_0$ to a good approximation. The resultant resistivities are
\begin{eqnarray}
\displaystyle{ \frac{\rho_{xx}^{\rm{col}}}{\rho_0}  =  \frac{ \Gamma }{4 \Gamma_0 } \left\{ 1 + 2\sum\limits_{k = 1}^\infty  {\left( 1 + 2\pi k \gamma \right)} \right. } \hspace{25mm} \nonumber \\
\displaystyle{ \Biggl. \times 
\cos \left[ 2\pi k\left( \varepsilon _\mathrm{F}  - \frac{1}{2} \right) \right]  J_0 ( 2\pi kv_B )e^{ - 2\pi k\gamma }  \Biggr\} }, \nonumber \\
\label{rhocolFin}
\end{eqnarray}
and
\begin{eqnarray}
\displaystyle{ \frac{\rho_{yy}^{\rm{dif}}}{\rho_0} = \frac{V_0 ^2}{ ak_\mathrm{F} E_\mathrm{F} \Gamma_0 } \omega_\mathrm{c} \tau \cos ^2 \left( qR_\mathrm{c}  - \frac{\pi }{4} \right) \Biggl\{ 1 +\Biggr.  } \hspace{15mm} \nonumber \\
\displaystyle{ \left.  4\sum\limits_{k = 1}^\infty  {\cos \left[ {2\pi k\left( \varepsilon _\mathrm{F}  - \frac{1}{2} \right)} \right]} \frac{J_1 (2\pi k v_B)}{2\pi k v_B}e^{ - 2\pi k\gamma }  \right\} }. \nonumber \\
\label{rhodifFin}
\end{eqnarray}
The $B$ dependence of eq.\ (\ref{rhocolFin}) inherits that of eq.\ (\ref{DOSULSL}), as expected, with minor discrepancy, to be discussed in the following subsection, resulting from the factor $(1+2 \pi k \gamma)$ in the summation. The first term in eq.\ (\ref{rhodifFin}) describes the CO and is exactly the same as the expression given in ref.\ \citen{Peeters92}. The amplitude of the CO increases linearly with $B$, as revealed by the factor $\omega_\mathrm{c} \tau$. (The damping factor due to scattering \cite{Endo00E} that results in additional $B$ dependence is not included here.) The second term represents the diffusion contribution to the SdHO\@. The amplitude of the oscillation is modulated by the CO, and is therefore enhanced (suppressed) at the maximum bandwidth (flat band) conditions; the phase of the modulation is at odds with that of the collisional contribution. Owing to the $B$-linear dependence of the CO mentioned above, the diffusion contribution to the SdHO raise its relative importance with $B$ and can outweigh the collisional contribution above a certain $B$. This qualitatively explains the observed transition, with the increase of $B$, from the suppression to the enhancement of the SdHO amplitude at the maximum bandwidth conditions. The interpretation will be confirmed in the next subsection by comparing the calculated traces with our experimental SdHO\@.

The numerical calculation of $\rho_{xx}^\mathrm{dif}$ by Shi \textit{et al}. \cite{Shi02} shows only negligibly small share of the SdHO\@. This can be interpreted in terms of the factor $J_1(2 \pi v_B) / 2 \pi v_B$ in eq.\ (\ref{rhodifFin}), keeping only the $k =$ 1 term in the summation. As shown in Fig.\ \ref{Bessels} (b), the function $J_1(2 \pi v_B) / 2 \pi v_B$ decrease from 0.5 with the increase of $|v_B|$, becomes zero at $|v_B| \simeq$ 5/8, and oscillates thereafter  with ever decreasing amplitude of less than 0.07, with zeros at $|v_B| \simeq 1/8+n/2$ ($n=2,3,4,...$). The oscillation of $v_B$ with $1/B$ works just to slightly counteract the effect of the CO while $|v_B|$ remains much smaller than 5/8, which is the case for our experiment for higher magnetic fields. When the amplitude of the periodic modulation is large, as is the case in Shi \textit{et al}., the factors $\cos^2 (q R_\mathrm{c} - \pi/4) $ and $|J_1(2 \pi v_B) / 2 \pi v_B|$ conspire to alternatingly become small and keep the diffusion contribution to the SdHO small over the whole range of magnetic fields.

\subsubsection{Comparison with experimental data}

\begin{figure}[t]
\includegraphics[bbllx=50,bblly=55,bburx=560,bbury=435,width=8.5cm]{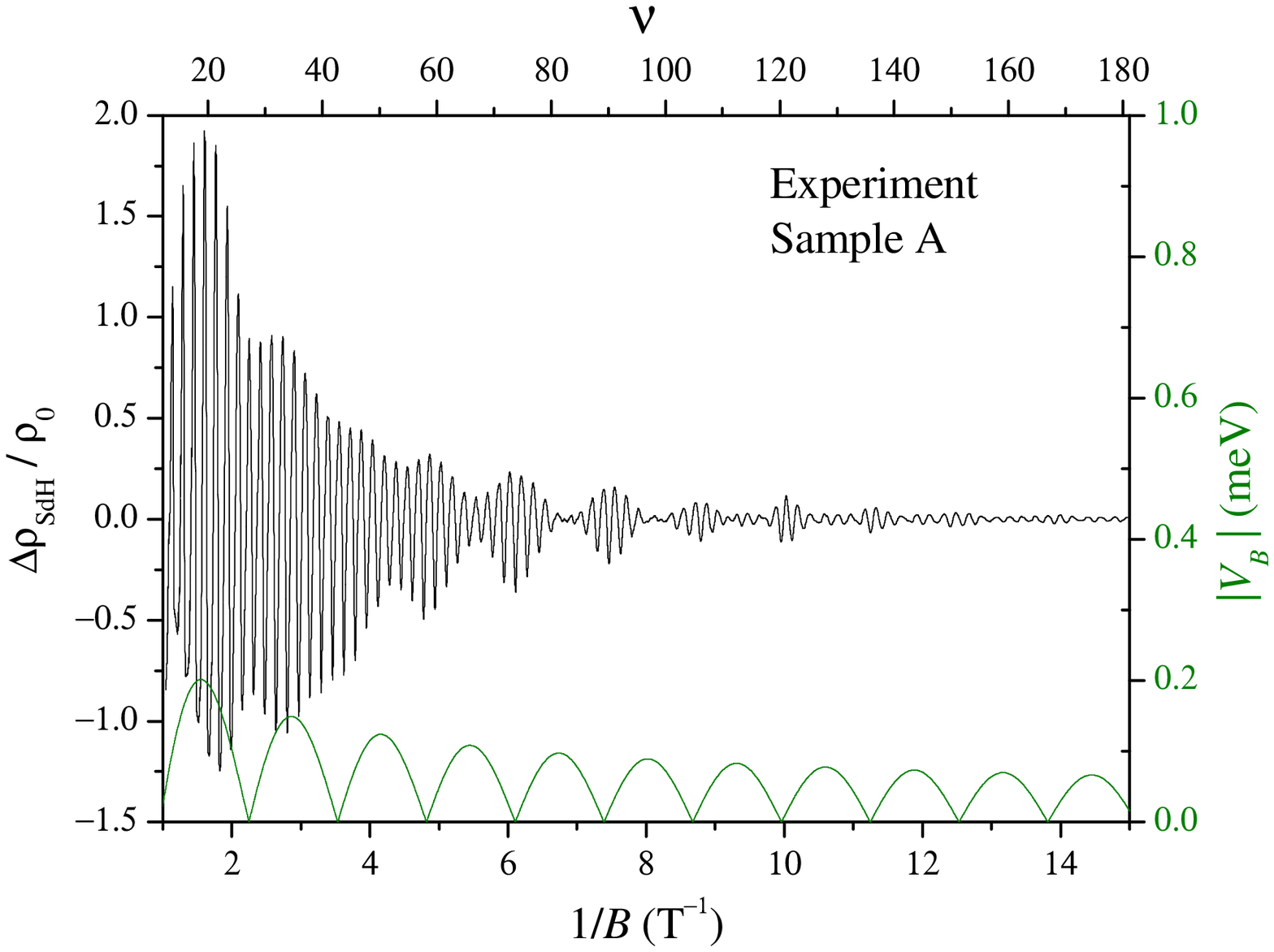}
\caption{(Color online) Experimentally obtained SdHO versus $1/B$ for sample A (left axis) and the half width of Landau bands (right axis).}
\label{Fexp}
\end{figure}

\begin{figure}[t]
\includegraphics[bbllx=50,bblly=100,bburx=560,bbury=780,width=8.5cm]{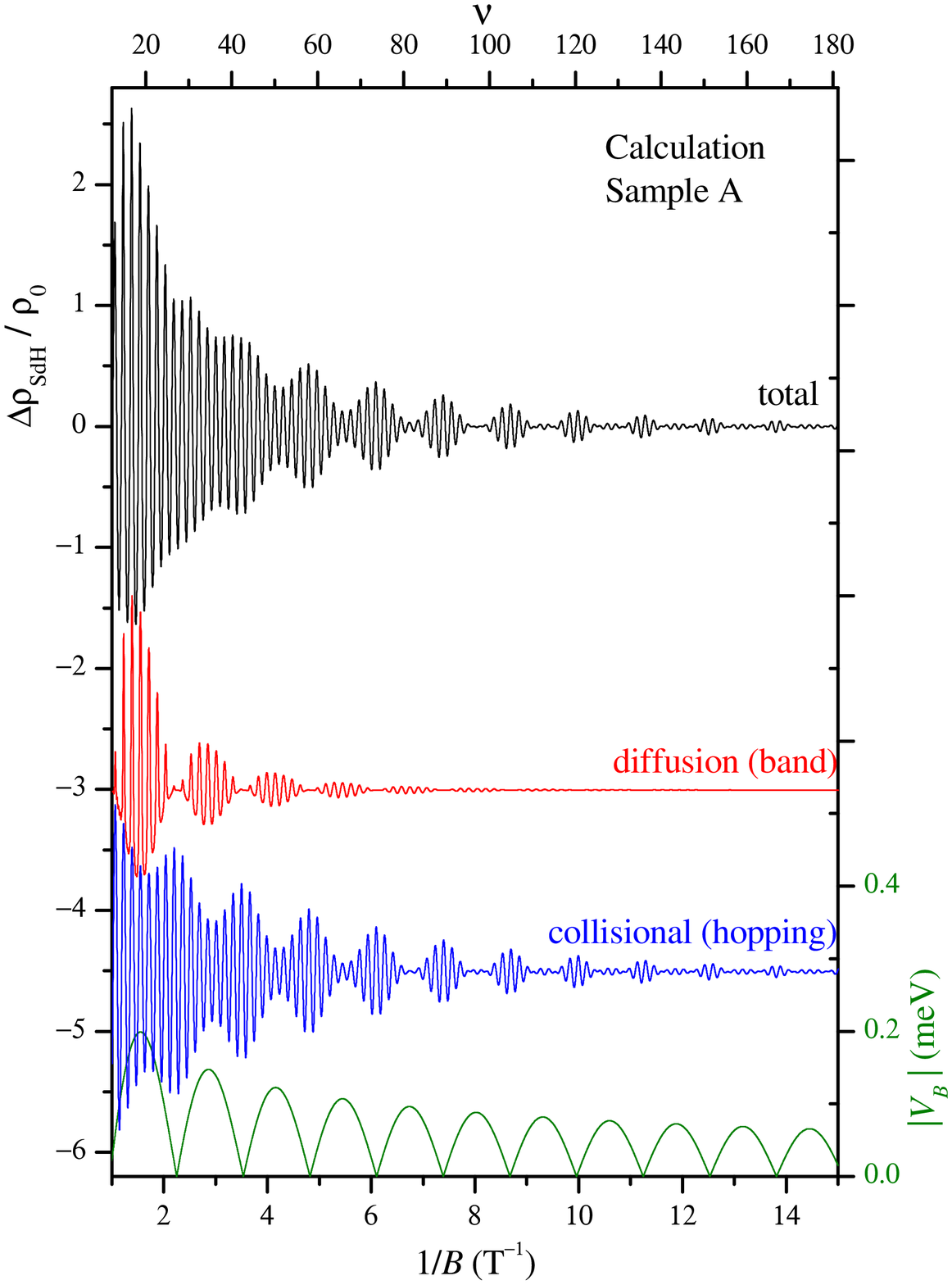}
\caption{(Color online) Calculated SdHO for Sample A. The collisional (hopping) contribution [eq.\ (\ref{rhoxxDOS})], the diffusion (band) contribution [eq.\ (\ref{rhodifA})], and the addition of the two contributions are plotted (left axis), with the former two traces negatively offset for clarity. The half width of the Landau bands is also plotted (right axis).}
\label{Fcalc}
\end{figure}

\begin{figure}[t]
\includegraphics[bbllx=50,bblly=55,bburx=560,bbury=435,width=8.5cm]{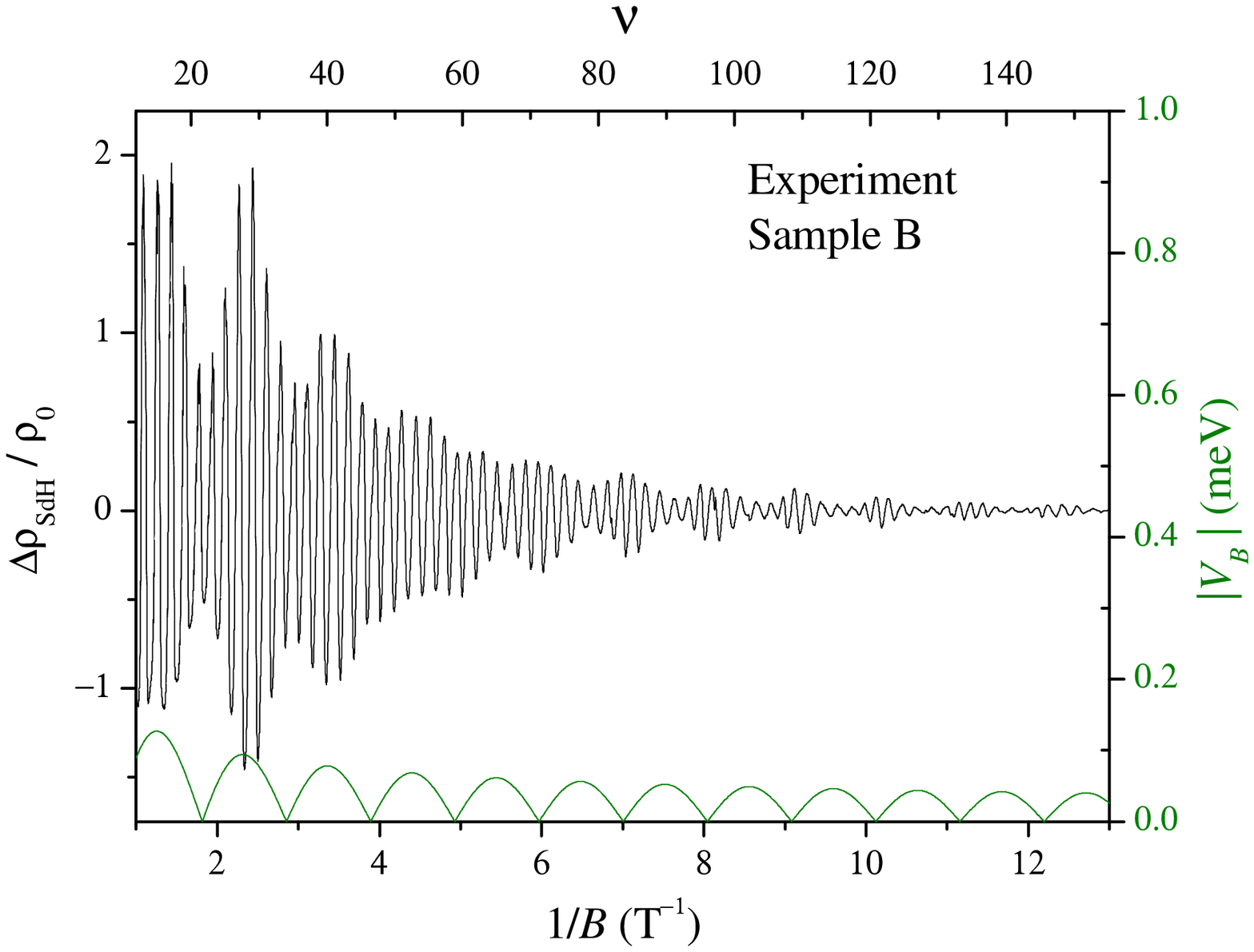}
\caption{(Color online) Experimentally obtained SdHO versus $1/B$ for sample B (left axis) and the half width of Landau bands (right axis).}
\label{Pexp}
\end{figure}

\begin{figure}[t]
\includegraphics[bbllx=50,bblly=100,bburx=560,bbury=780,width=8.5cm]{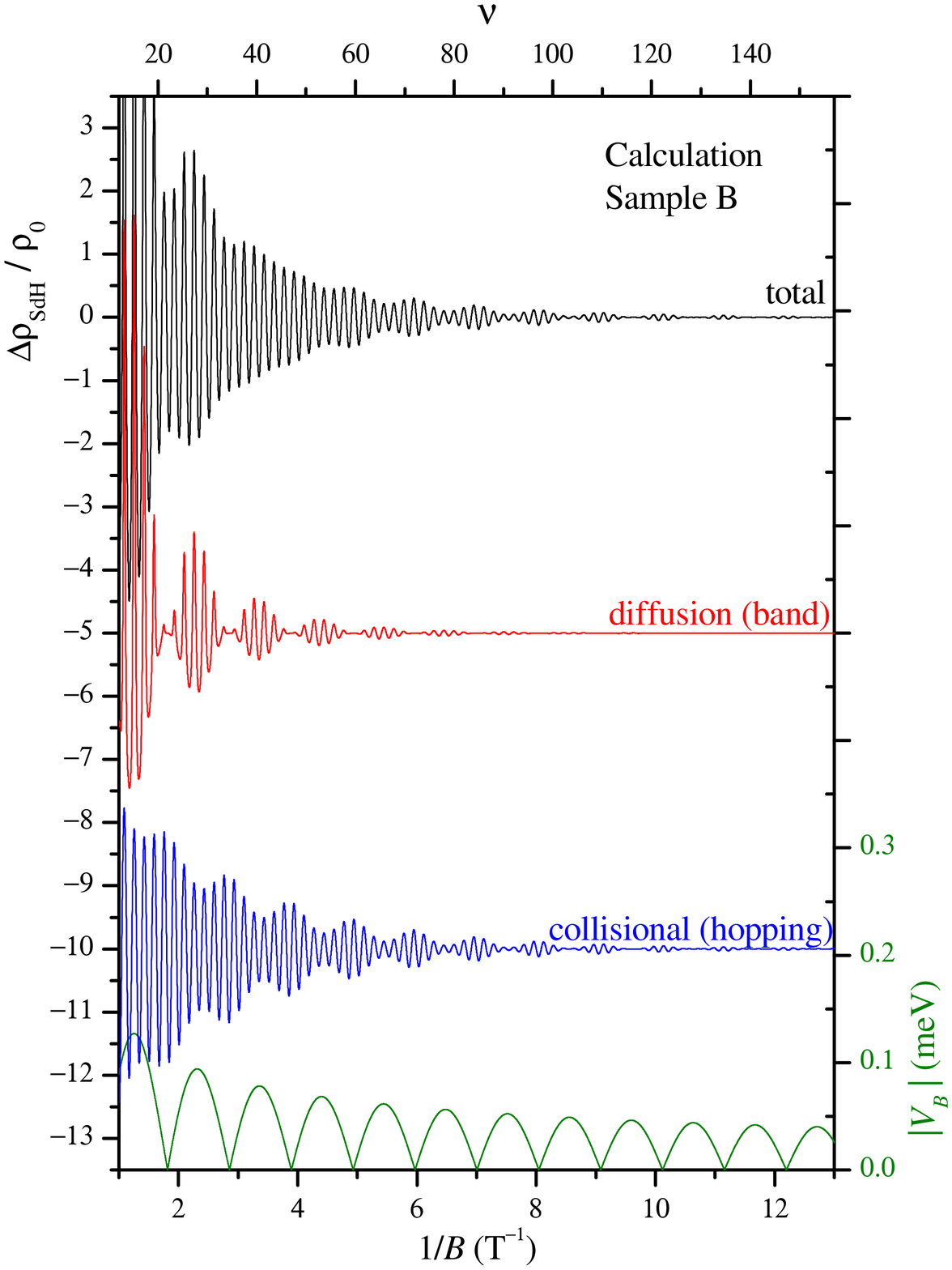}
\caption{(Color online) Similar to Fig.\ \ref{Fcalc} for Sample B\@.}
\label{Pcalc}
\end{figure}

In this subsection, we compare our experimentally obtained SdHO with calculated resistivities, in an attempt to gain more quantitative understanding of the magnetic-field dependence of the SdHO amplitude. Main focus is on the behavior at higher magnetic-field side that remains unexplained in \S \ref{cmpDOS}.

Figures \ref{Fexp} and \ref{Pexp} show the experimentally obtained SdHO for Sample A and Sample B, respectively, plotted against $1/B$. The oscillatory parts, $\Delta \rho_\mathrm{SdH} / \rho_0$, are obtained by applying a Fourier high-pass filter to the plot of $\rho_{xx} / \rho_0$ vs. $1/B$. Simultaneously plotted $|V_B|$ serves as a guide to review the transition with the increase of $B$ from suppression to enhancement of the SdHO amplitudes at the maxima of the band width. 

In Figs.\ \ref{Fcalc} and \ref{Pcalc}, we plot calculated collisional (hopping) and diffusion (band) contributions and the sum of the two contributions for Sample A and Sample B, respectively, using corresponding sample parameters. In pursuit of better agreement with the experiment, we used slightly modified version of the argument presented in the previous subsection.

As mentioned earlier, the collisional contribution given by eq.\ (\ref{rhocolFin}) possesses an additional $B$-dependent factor $(1 + 2 \pi k \gamma)$ compared with the DOS in eq.\ (\ref{DOSULSL}). In principle, eq.\ (\ref{rhocolFin}) should also describe the SdHO of the plain 2DEG by placing $V_0 =$0 ($v_B \equiv$ 0). This, however, is at variance with the firmly established relation $\Delta \rho_\mathrm{SdH} / [\rho_0 A(T/T_\mathrm{c})] \propto \Delta D_1 / D_0$\cite{Coleridge89} owing to the extra factor, which may possibly be resulting from the approximation used in the course of deducing eq.\ (\ref{sgmcol}). We, therefore, discard eq.\ (\ref{rhocolFin}) and assume that the relation
\begin{equation}
\frac{\Delta \rho_{xx}^\mathrm{col}}{\rho_0} = \frac{\rho_{xx}^\mathrm{col} - \rho_0}{\rho_0} \simeq C \frac{\Delta D_1}{D_0}
\label{rhoxxDOS}
\end{equation}
confirmed for plain 2DEGs also represents the oscillatory part of the collisional contribution of the ULSLs. Note, however, that this choice does not have a drastic effect, since $(1 + 2 \pi k \gamma) \simeq 1$ for large enough magnetic fields. In eq.\ (\ref{rhoxxDOS}), we assumed $A(T/T_\mathrm{c}) \simeq$1 appropriate for low temperatures. The constant $C$ has been shown to be equal to 2 for ideally uniform 2DEGs but deviate from this ideal value by a small (typically a few percent) inhomogeneity in the electron density \cite{Coleridge91}. We selected $C =$ 0.92 and 2 for Sample A and Sample B, respectively, the values that quantitatively describe the SdHO of the adjacent plain 2DEGs and also the SdHO of the ULSLs for the low magnetic-field range \cite{prefactor}.

In a previous publication \cite{Endo00E}, we have pointed out that a factor $A(\pi / \omega_\mathrm{c} \tau_\mathrm{w})$ describing the damping of the CO due to scattering needs to be incorporated for quantitative account of the experimental CO traces, with $\tau_\mathrm{w}$ a characteristic scattering time usually identifiable with $\tau_\mathrm{Q}$. This will also affect the second term in eq.\ (\ref{rhodifFin}) that includes the CO as a multiplying factor, resulting in the diffusion contribution to the SdHO (excluding the first term corresponding to the ordinary CO) as
\begin{eqnarray}
\displaystyle{ \frac{\Delta \rho_{yy}^{\rm{dif}}}{\rho_0} = A\left( \frac{\pi}{\omega_\mathrm{c} \tau_\mathrm{w}} \right) \frac{4 V_0 ^2 \omega_\mathrm{c} \tau}{ ak_\mathrm{F} E_\mathrm{F} \Gamma_0 } \cos ^2 \left( qR_\mathrm{c}  - \frac{\pi }{4} \right)   } \hspace{15mm} \nonumber \\
\displaystyle{ \times \sum\limits_{k = 1}^\infty  {\cos \left[ {2\pi k\left( \varepsilon _\mathrm{F}  - \frac{1}{2} \right)} \right]} \frac{J_1 (2\pi k v_B)}{2\pi k v_B}e^{ - 2\pi k\gamma }   }. \nonumber \\
\label{rhodifA}
\end{eqnarray}
Thermal damping is neglected here again. The exponential factor $\exp (-2 \pi k \gamma)$ still works in favor of smaller $k$. However, due to the extra $B$-linear dependence mentioned earlier, the more weight is on the higher magnetic field side for the diffusion contribution, where the exponential factors still remain rather large. Therefore, in Figs.\ \ref{Fcalc} and \ref{Pcalc}, we preserved up to $k =$ 5 in the summation of eq.\ (\ref{rhodifA}).

By comparing Figs.\ \ref{Fexp}, \ref{Pexp} and Figs.\ \ref{Fcalc}, \ref{Pcalc}, it can be seen that the addition of the two types of contributions qualitatively reproduce the experimental SdHO, notably the transition from the suppression to the enhancement of SdHO at maximum bandwidth conditions. The transition is attributable to the rapid growth of the diffusion contribution with increasing $B$. The diffusion contribution plays more important role in Sample B than in Sample A\@. This can be ascribed to the higher mobility $\mu$ for Sample B; as can be seen in eq.\ (\ref{rhodifA}), the diffusion contribution is proportional to $\tau / \Gamma_0 \propto \mu^2$. The effect of smaller modulation amplitude $V_0$ is two fold: on one hand, the smaller $V_0$ makes the counteracting effect of $J_1(2 \pi v_B)/(2 \pi v_B)$ mentioned at the end of the preceding subsection less effective; on the other hand, the smaller $V_0$ is disadvantageous because of the factor ${V_0}^2$ in eq.\ (\ref{rhodifA}). These two effects somewhat compensate each other to make the difference in $V_0$ between the two samples less important.

The degree of agreement between the calculated and the experimental SdHO shown here should be assessed with care, since we had to employ slightly larger values of $V_0$ in eq.\ (\ref{rhoxxDOS}) than in eq.\ (\ref{rhodifA}) in order to achieve good agreement with experimental traces, as mentioned in \S \ref{cmpDOS}. Although the inconsistency should be resolved in the future studies, it does not affect the qualitative argument presented here. 

\section{Conclusions}
We have shown that the experimentally observed SdHO for a ULSL is basically reproduced by the addition of the collisional contribution $\Delta \rho_{xx}^\mathrm{col}$ [eq.\ (\ref{rhoxxDOS}) with eq.\ (\ref{D1ULSL})] and the diffusion contribution $\Delta \rho_{xx}^\mathrm{dif}$ [eq.\ (\ref{rhodifA})]. The amplitude of the oscillation alternates  between the two contributions: $\Delta \rho_{xx}^\mathrm{col}$ is suppressed while $\Delta \rho_{xx}^\mathrm{dif}$ is enhanced at the maximum bandwidth conditions [eq.\ (\ref{bandmax})]. Owing to an extra linear-$B$ factor for $\Delta \rho_{xx}^\mathrm{dif}$ in addition to the common exponential damping factor, $\Delta \rho_{xx}^\mathrm{col}$ dominates the SdHO at low magnetic fields but $\Delta \rho_{xx}^\mathrm{dif}$ outweighs $\Delta \rho_{xx}^\mathrm{col}$ at higher magnetic fields. This accounts for the experimentally observed transition from suppression to enhancement of the SdHO at the maximum bandwidth conditions. The term $J_1(2 \pi v_B)/(2 \pi v_B)$ in eq.\ (\ref{rhodifA}) qualitatively explains why diffusion contribution to the SdHO was not observed in a previous experiment \cite{Edmonds01}.

\section*{Acknowledgment}
This work was supported by Grant-in-Aid for Scientific Research (C) (18540312) and (A) (18204029) from the Ministry of Education, Culture, Sports, Science and Technology (MEXT).

\appendix
\section{Derivation of eq.\ (\ref{sgmcolFin})}
In this appendix, we describe the derivation of eq.\ (\ref{sgmcolFin}) from eq.\ (\ref{sgmcolLT}). Performing the summation first, eq.\ (\ref{sgmcolLT}) becomes
\begin{equation}
\sigma_{xx}^{\rm{col}}  = \frac{e^2}{h}\Gamma ^2 \pi \frac{1}{\pi }\int_0^\pi  {dt S(E_\mathrm{F},t)}
\label{sgmxxcolS}
\end{equation}
with
\begin{eqnarray}
\displaystyle{ S(E_\mathrm{F},t) }& = &\displaystyle{ \sum\limits_{N = 0}^\infty  (2N + 1) P^2 (E_\mathrm{F}  - E_{N,t})}  \nonumber \\
 & = & \displaystyle{ \left( \frac{\gamma}{\pi \hbar \omega_\mathrm{c}} \right)^2 \sum\limits_{N = 0}^\infty \frac{2 N + 1}{\left[ (N-\alpha)^2+\gamma^2 \right] ^2} } \nonumber \\
 & = & \displaystyle{ \left( \frac{\gamma}{\pi \hbar \omega_\mathrm{c}} \right)^2 \left( -\frac{1}{2 \gamma} \right) \frac{\partial }{\partial \gamma} \sum\limits_{N = 0}^\infty \frac{2 N + 1}{(N-\alpha)^2+\gamma^2} }, \nonumber \\
\label{Sum}
\end{eqnarray}
where we introduced the notation $\alpha = \varepsilon_\mathrm{F}  - 1/2- v_B \cos t$ for brevity. Noting that $2 N+1 = 2 \alpha+1+2(N-\alpha)$, the summation in the last line of eq.\ (\ref{Sum}) can be performed by appealing to the approximation $\sum_{N=0}^\infty$ $\rightarrow$ $\sum_{N=-\infty}^\infty$ as before, and by applying the Poisson sum formulae, resulting in
\begin{eqnarray}
\displaystyle{ \pi \left\{ \frac{2 \alpha +1}{\gamma } \left[ 1 + 2 \sum\limits_{k = 1}^\infty { \cos (2 \pi k \alpha ) e^{-2 \pi k \gamma} } \right] \right.} \nonumber \\
\displaystyle{ \left. -4 \sum\limits_{k = 1}^\infty { \sin (2 \pi k \alpha) e^{-2 \pi k \gamma} } \right\} }.
\end{eqnarray}
Replacing this result into eq.\ (\ref{Sum}), we obtain
\begin{eqnarray}
\displaystyle{ S(E_\mathrm{F},t) = \frac{1}{( \hbar \omega_\mathrm{c} )^2}} \hspace{55mm} \nonumber \\
\displaystyle{ \times \left\{ \frac{2 \alpha +1}{2 \pi \gamma} \left[ 1 + 2 \sum\limits_{k = 1}^\infty { (1 + 2 \pi k \gamma ) \cos ( 2 \pi k \alpha ) e^{-2 \pi k \gamma} } \right] \right. } \nonumber \\
\displaystyle{ \left. -4 \gamma \sum\limits_{k = 1}^\infty {k \sin (2 \pi k \alpha) e^{-2 \pi k \gamma}} \right\} }. \nonumber \\
\end{eqnarray}
Performing the integral in eq.\ (\ref{sgmxxcolS}), we finally achieve,
\begin{eqnarray}
\displaystyle{ \frac{\sigma_{xx}^{\rm{col}}}{\sigma_0} }  & = & \displaystyle{ \frac{\Gamma _0}{\Gamma }\gamma ^2 \left\{ 1 + 2\sum\limits_{k = 1}^\infty  {\left( 1 + 2\pi k \gamma \right)} \right. } \hspace{25mm} \nonumber \\
 & & \hspace{5mm} \displaystyle{ \Biggl. \times 
\cos \left[ 2\pi k\left( \varepsilon _\mathrm{F}  - \frac{1}{2} \right) \right]  J_0 ( 2\pi kv_B )e^{ - 2\pi k\gamma }  \Biggr\} } \nonumber \\
 & & \displaystyle{- 4\pi \frac{{\Gamma _0 }}{{E_\mathrm{F} }}\gamma ^3 \sum\limits_{k = 1}^\infty  {k\sin \left[ {2\pi k\left( {\varepsilon _\mathrm{F}  - \frac{1}{2}} \right)} \right]} } \nonumber \\
 & & \hspace{5mm} \displaystyle{ \times J_0 \left( {2\pi kv_B } \right)e^{ - 2\pi k\gamma } } \nonumber \\
 & & \displaystyle{ - \frac{\Gamma _0 }{\Gamma}\gamma ^2 \frac{V_B }{E_\mathrm{F}}2\sum\limits_{k = 1}^\infty  {\left( 1 + 2\pi k \gamma \right)} \hspace{15mm}}  \nonumber \\
 & & \hspace{5mm} \displaystyle{ \times \sin \left[ 2\pi k\left( \varepsilon _\mathrm{F}  - \frac{1}{2} \right) \right] J_1 (2\pi kv_B)e^{ - 2\pi k\gamma } }.\nonumber \\
\end{eqnarray}
Owing to the smallness of $\Gamma_0 / E_\mathrm{F}$ and $V_B / E_\mathrm{F}$, the first term is by far the dominant term, ending up with eq.\ (\ref{sgmcolFin}).

\bibliography{twodeg,lsls,magmod,ourpps,noteModSdH}

\begin{thebibliography}{10}
\bibitem{NoteDiffCryo}
The trace at 4.2 K was taken with a different cryostat from other lower
  temperature traces, and is slightly rescaled, both horizontally and
  vertically, to correct for a small difference in the electron density.
\bibitem{highharm}
Peaks $f_\mathrm{SdH} + 2 f_\mathrm{CO}$ and $f_\mathrm{SdH} + 3 f_\mathrm{CO}$
  are also observed for Sample B at a different electron density,
  $n_e$=2.11$\times$10$^{15}$ m$^{-2}$.
\bibitem{approxminus}
Here, we made an approximation, $\sum_{N=0}^\infty$ $\rightarrow$
  $\sum_{N=-\infty}^\infty$; the approximation exerts negligible effect on
  $D(E)$ for $\Gamma \ll E$.
\bibitem{constEF}
Here, the $E_\mathrm{F}$ is approximated by a constant value at $B =$ 0.
  Strictly speaking, $E_\mathrm{F}$ oscillates with $B$ to fulfill $n_e =
  \int_0^{\infty } f(E) D(E) \, dE$, with $f(E)$ the Fermi-Dirac distribution
  function. However the effect is small at low magnetic fields and is
  negligible in the context of the present study.
\bibitem{stilllarge}
The function $J_0(2 \pi v_B)$ oscillates with $v_B$ for still larger $|v_B|$,
  alternating the sign at $|v_B| \simeq (n+3/4)/2$ ($n$=0,1,2,...). The
  oscillation of $\Delta D_1$ switches the peak/dip positions accordingly. The
  modulation amplitudes $V_0$ for our samples are too small to reach such
  regime.
\bibitem{evenodd}
Similar relations lead to the occurrence of the even-odd transition at $2 |V_B|
  \simeq (n+3/4) \hbar \omega_\mathrm{c}$ ($n$=0,1,2,...), which is slightly
  shifted from the corresponding conditions $2 |V_B| = (n+1) \hbar
  \omega_\mathrm{c}$ ($n$=0,1,2,...) proposed in ref.\ \citen{Shi02}.
\bibitem{difdef}
Note that the definition of $\sigma_0$ here is different from that in ref.\
  \citen{Shi02}.
\bibitem{prefactor}
The prefactor in eq.\ (\ref{rhocolFin}) implies that the constant $C$ might
  depend on the scattering parameters, in contradiction to the interpretation
  in ref.\ \citen{Coleridge91} that attributes the deviation of $C$ from 2 to
  the non-uniformity of the 2DEG\@. Substituting our sample parameters, we
  obtain the prefactor $\Gamma / 4 \Gamma_0 =$ 1.45 and 3.47 for Sample A and
  Sample B, respectively. The ratio of the prefactor between the two samples
  roughly preserves that of $C$ we have selected. It requires further studies
  to clarify whether this actually signifies the involvement of the scattering
  parameters to the value of $C$ or it is just a coincidence.
\end{thebibliography}

\end{document}